\documentclass[useAMS, usenatbib, usegraphicx]{mn2e}
\usepackage{amssymb}
\usepackage{booktabs}
\newcommand\ApJ{ApJ}
\newcommand\MNRAS{MNRAS}

\newcommand\AnA{A\&A}

\def\alf{Alfv\'en\,}

\def\sc{Sch$\ddot{\mbox{u}}$ssler\,}
\def\as{\'a}
\def\es{\'e}
\def\isp{\'i}
\def\bq{\begin{equation}}
\def\eq{\end{equation}}
\def\ee #1 {\times 10^{#1}}
\def\ut #1 #2 { \, \rmn{#1}^{#2}}
\def\u #1 { \, \rmn{#1}}

\def\kms {\,\rmn{km\,s}^{-1}}
\def\persec {\, \hbox{s}^{-1}}

\let\grad=\nabla
\newcommand\cross{\bmath{\times}}

\newcommand\etaH{\eta_\mathrm{H}}
\newcommand\etaP{\eta_\mathrm{P}}

\newcommand\hH{\tilde{\eta}_\mathrm{H}}
\newcommand\hP{\tilde{\eta}_\mathrm{P}}
\newcommand\hA{\tilde{\eta}_\mathrm{A}}
\newcommand\hO{\tilde{\eta}_\mathrm{O}}
\newcommand\hT{\tilde{\eta}_\mathrm{T}}
\newcommand\het{\tilde{\eta}}

\def\curl{{\grad \cross}}
\def\div #1 {\grad \cdot #1}
\def\vp{v_0^\prime}
\def\kz{\hat{k_z}}

\def\hk{\tilde{k}}
\def\hs{\tilde{\sigma}}

\def\v{\bmath{v}}
\def\dv{\bmath{\delta\v}}
\def\hdv{\bmath{\delta\hat{\v}}}

\def\dvx{\delta v_x}
\def\dvy{\delta v_y}
\def\dvz{\delta v_z}

\def\k{\bmath{k}}

\def\vi{\bmath{v}_i}

\def\ve{\bmath{v}_e}
\def\vi{\bmath{v}_i}

\def\vn{\bmath{v}_n}
\def\vB{\bmath{v}_B}
\def\dvB{\bmath{\delta\vB}}

\def\J{\bmath{J}}
\def\B{\bmath{B}}
\def\dBx{\delta B_x}
\def\dBy{\delta B_y}
\def\dBz{\delta B_z}

\def\E{\bmath{E}}            % E
\def\hB{\bmath{b}}
\def\kz{\hat{k}_z}
\def\kx{\hat{k}_x}
\def\Bz{b_z}
\def\By{b_y}
\def\avp{|\vp|}

\def\hdvx{\hat{\delta{v_x}}}

\def\J{\bmath{J}}
\def\Jpa{\bmath{J_\parallel}}  % J_||
\def\Jpe{\bmath{J_\perp}}  % J_perp

\def\dB{\bmath{\delta\B}}
\def\hdB{\bmath{\delta\hat{\B}}}

\newcommand{\delt} [1] {\frac{\partial #1}{\partial t}}

\title{Magnetic diffusion driven shear instability of solar flux tubes}
\author[B.P.Pandey and Mark Wardle]
        {B.P. Pandey and Mark Wardle \\
{Department of Physics, Astronomy \& Research Centre for Astronomy, Astrophysics \& Astrophotonics,}\\Macquarie University, Sydney, NSW 2109, Australia}
\date{\today}
\pagerange{\pageref{firstpage}--\pageref{lastpage}}
\pubyear{2012}
\begin{document}
\maketitle
\label{firstpage} 
\begin{abstract}
\maketitle 
The dynamics of the partially ionised solar photosphere and chromosphere can be described by a set of equations that are structurally similar to the magnetohydrodynamic equations, except now the magnetic field is no longer \emph{frozen} in the fluid but slips through it due to non--ideal magnetohydrodynamic effects which are manifested as Ohm, ambipolar and Hall diffusion. Macroscopic gas motions are widespread throughout the solar atmosphere and shearing motions couple to the non--ideal effects, destabilising low frequency fluctuations in the medium. The origin of this non-ideal magnetohydrodynamic instability lies in the collisional coupling of the neutral particles to the magnetized plasma in the presence of a sheared background flow. Unsurprisingly, the maximum growth rate and most unstable wavenumber depend on the flow gradient and ambient diffusivities.  

The orientation of the magnetic field, velocity shears and perturbation wave vector play a crucial role in assisting the instability. When the magnetic field and wave vector are both vertical, ambipolar and Ohm diffusion can be combined as Pedersen diffusion and cause only damping; in this case only Hall drift in tandem with shear flow drives the instability.  However, for non-vertical fields and oblique wave vectors, both ambipolar diffusion and Hall drift are destabilizing.

We investigate the stability of magnetic elements in the network and internetwork regions. The shear scale is not yet observationally determined, but assuming a typical shear flow gradient $\sim 0.1 \,\mbox{s}^{-1}$ we show that the magnetic diffusion shear instability grows on a time scale of one minute.  Thus, it is plausible that network--internetwork magnetic elements are subject to this fast growing, diffusive shear instability, which could play an important role in driving low frequency turbulence in the plasma in the solar photosphere and chromosphere.   
\end{abstract}


\section{Introduction}
The solar atmosphere is host to a variety of highly energetic events such as coronal mass ejections (CMEs), flares, prominences, and coronal heating. The huge reservoir of energy in the solar atmosphere is related to the magnetic field originating in the convection zone, where shearing motion of field line foot points can stretch and twist the anchored field \citep{P79}. A variety of magnetohydrodynamic (MHD) waves are generated in the corona due to convective motion in the photosphere. The convective shearing motion not only produces \alf, fast or, slow magnetoacoustic waves, but also brings topologically disparate parts of the magnetic configuration closer together resulting in the formation of current sheets. All in all, a tiny fraction of convective energy carried by the waves to higher altitude may suffice to heat the coronal plasma to high temperatures. This simple physical picture of wave excitation, propagation and the ensuing coronal heating is attractive as it ties the heat transport in the solar corona and heliosphere to the ultimate source of energy -- shearing photospheric convective motions. Therefore the investigation of wave propagation and concomitant heating of coronal plasma in the framework of MHD has been a popular topic in solar physics \citep{P87, P97a,  P97b, P98, H99, GP04, A06}. 

Most solar atmospheric heating, with the possible exception of flares, takes place in the chromosphere. The chromosphere extends up to about nine pressure scale heights, i.e. about $2$\,Mm above the photospheric surface.  The lower chromosphere is threaded by strong ($\sim$ kiloGauss) vertical flux tubes located in the network regions where they are observed as bright points. These tubes, which have a low filling factor ($< 1 \%$) near the foot point in the photosphere, expand with increasing height to fill about $15 \%$ of the lower chromosphere (altitude $\sim 1 \mbox{Mm}$, where CaII emission features are observed in H and K lines) before filling the entire atmosphere and forming a canopy in the chromosphere. The quiet solar internetwork region is also magnetised, with patches of concentrated kG magnetic field \citep{D09, SA10} and an order of magnitude smaller field everywhere else. Observation suggests that localised active regions that emit $\sim 80 \%$ of the coronal radiative loss near solar maximum contain plasma of chromospheric origin \citep{A01}. This raises the possibility that the same mechanism that transports mechanical energy from the convection zone to the chromosphere to sustain its heating rate also supplies the energy needed to heat the corona and accelerate the solar wind. 

However, the number of plasma particles in the partially ionised solar atmosphere, particularly below   $\lesssim 2.5 \mbox{Mm}$ is small \citep{VAL81}. As a result the plasma particles are not {\it frozen} in the field owing to frequent collision with the neutral hydrogen \citep{MS56}. Therefore, the ideal MHD description which is valid only for the fully ionised fluid is unsuitable to describe the low temperature photosphere-chromosphere where non--ideal MHD effects such as Hall, ambipolar and Ohm diffusion dominate \citep{PW06, P08a, P08b, K12, SK12, PW12a}. All in all, there is no unified ideal MHD like framework to describe the weakly ionised and weakly magnetised photosphere and fully ionised and highly magnetised corona with weakly ionised and highly magnetised chromosphere sandwiched in-between. 

The inclusion of neutral dynamics not only destroys the economy and simplicity of the single fluid MHD description of fully ionised plasma but the very concept of a flux tube may be difficult to define in the multi--fluid framework.  Furthermore, high frequency \alf waves may not survive the collision-dominated photosphere and chromosphere \citep{G04, LAK05, VPP07, AHL07, VPPD07, G11}. This could have been anticipated on the grounds that in the photosphere ($\lesssim 500 \,\mbox{km}$) and chromosphere ($\lesssim 2500 \,\mbox{km}$) the plasma number density  is  much smaller than the neutral number density and thus, high frequency (with respect to the ion-neutral collision frequency) MHD waves are severely damped by collisional dissipation of the wave energy. A way out of this difficulty is to retain the MHD momentum equation for the ionized component and include the effect of collisions with the neutrals by modifying the induction and energy equations to incorporate a conductivity tensor \citep{E04, L06}, an approach often employed to study the lower ionosphere of the Earth. However, the derivation of the \emph{time--independent} conductivity tensor neglects time derivatives in the electron and ion momentum equations, i.e. $d_{e\,,i}/dt \sim \omega_{e, i} \ll \omega_{ce, ci}$, requiring that the dynamical response frequencies of the ions and electrons, $\omega_{e, i}$, are much smaller than their respective gyro-frequencies, $\omega_{ce, ci}$. Further, the relative ion--neutral drift is assumed much smaller than the centre of mass ion--neutral velocity \citep{MK73}.  Therefore, neglecting plasma inertia, a linear relationship between the electric field $\E$ and plasma current $\J$ can be easily derived $\E = \bmath{\sigma}\cdot \J$ where $\bmath{\sigma}$ is the \emph{time-independent} conductivity tensor. However, the MHD equation of motion assumes $\omega_{i} \sim \omega_{ci}$. In fact the ion carries the inertia of the fluid. Therefore, on the one hand, the \emph{time--independent} conductivity tensor implies $\omega_{e, i} \ll \omega_{ce, ci}$, on the other hand, the MHD momentum equation requires $\omega_{i} \sim \omega_{ci}$. Clearly, investigation of the collisional effects by merely modifying the induction and energy equation in the MHD framework is highly unsatisfactory.
 
Any realistic model of the solar atmosphere must reflect two basic observational facts: (a) the magnetic field distribution on the solar surface is not continuous but is organised into network and internetwork elements. Whereas the network field ($\gtrsim \mbox{kG}$) is predominantly vertical and organised into flux tubes (diameter $\lesssim 100\,\mbox{km}$) located in the intergranular lanes, the internetwork field $(\mbox{few}\,\mbox{G}-\mbox{kG})$ in the interior of supergranule cells is primarily horizontal \citep{H09, L08} \footnote{The internetwork field may instead be isotropic \citep{SA11}, but see also \cite{S12}.}; (b) the plasma in the photosphere--chromosphere is weakly ionised (with fractional ionisation, i.e.\ the ratio of the electron and neutral number densities, $X_e = n_e / n_n \sim 10^{-4}$, VAL81).

Both the active and quiet phases of the solar atmosphere are highly dynamic and consist of convectively driven vortices and flows on a variety of spatial and temporal scales \citep{B08, W09, Ba10, B10}. Most vortices are small ($\lesssim 0.5\,\mbox{Mm}$) with average size $\sim 241 \pm 25\,\mbox{km}$ and typical lifetime $\sim 3-5\,\mbox{min}$, although large vortices $\sim 20\,\mbox{Mm}$ with lifetime $\gtrsim 20\,\mbox{min}$ have also been observed \citep{At09}.  The bright points associated with the vortex motion in the intergranular lane moves with typical speed $\lesssim 2\,\mbox{km} / \mbox{s}$ \citep{W09}. Magnetic fields appear to play a crucial role in mediating vortex motion in the photosphere and chromosphere \citep{S12}.

Rotation has been invoked in the past to explain the stability of flux tubes \citep{S84}. Models of spicules also invoke rotating flux tubes \citep{KS97}. Numerical simulations of solar convection display turbulent vortex flows at intergranular lanes \citep{Z93, SN98}. Vorticity generation near the boundaries of granules has also been seen in numerical simulation of the photosphere \citep{N09, M10}. The formation of small-scale, intergranular vortices suggests that vorticity is formed due to the interaction of photospheric plasma with the ambient magnetic field in intergranular lanes \citep{M11, S11}.  High-resolution simulations including the effects of non-ideal MHD show that the Hall effect generates out-of-plane velocity fields with maximum speed $\sim 0.1\,\mbox{km} / \mbox{s}$ at the interface layers between weakly magnetized light bridges and neighbouring strong field umbral regions \citep{C12}. To summarise, both observation and numerical simulation points to the presence of shear flow at various spatial scales in the solar photosphere. 

Large scale shear flow acts as a source of free energy in the solar plasma that can easily destabilise waves. For example, shear driven Kelvin-Helmholtz instability (KHI), which converts shear flow energy into vortex kinetic energy, is invoked to explain the dynamical structures in the solar atmosphere \citep{ J93, Kl04, Sr10, Z10, OF11, F11}. The KHI possibly acts as a triggering mechanism for large scale solar transient phenomena such as solar flares, CMEs, and associated eruptions \citep{Sva12}. The generation of the highly dynamical structures observed by the Solar Dynamical Observatory (SDO) is most likely due to the KHI \citep{F11}. 

Although the solar atmosphere may be susceptible to KHI, the presence of a magnetic field is not conducive to this instability. For example, a magnetic field directed along the flow suppresses KHI whereas a transverse field has no effect on the instability \citep{C61}. However, magnetic fields not only quench shear instabilities but can also facilitate them. For example, the most important instability in accretion discs, the magnetorotational instability is caused by an interplay between the angular velocity of the magnetised fluid and magnetic field \citep{BH98}. Thus, depending on the presence of a velocity gradient, magnetic field, when well coupled to the plasma can suppress as well as drive the instability. Therefore, before dwelling upon the role of the magnetic field on the flow driven instabilities, it is pertinent to know how well is the magnetic field coupled to the surrounding matter.     

Magnetic field drift through weakly ionised matter in the presence of shear flow can assist waves to grow.  For example, in protoplanetary discs, both Hall and ambipolar diffusion enhance the magnetorotational instability \citep{W99, BT01, KB04, D04, WS12, PW12b}. Clearly, the drift of the magnetic field in a weakly-ionised diffusive medium provides new pathways through which shear energy can be channelled to the waves. Indeed, diffusive destabilisation of a partially ionised medium in the presence of shear flow is not unique to weakly-ionised discs but is generic \citep{K08, PW12a}. The crucial ingredients required to excite this diffusion--shear instability are the presence of a shear flow, and favourable magnetic field topology. The ensuing instability is overtly similar to KHI, and not surprisingly, the growth rate is proportional to the shear gradient. However, unlike KHI which is hydrodynamic in nature, this is a magnetohydrodynamic instability.

A detailed investigation of diffusive shear instability in the context of the solar atmosphere is carried out here, building on our previous work (\citealt{PW12a}, hereafter PW12a), as follows.  First, unlike PW12a, where only a vertical field and transverse fluctuation (vertical wave vector) is assumed, here field topology is more general and the wave vector may be oblique.  Second, in PW12a the back reaction of the fluid on the magnetic field was completely ignored, whereas here it is retained and as a result the shear driven diffusive instability does not have a cut-off wavelength. Third, for vertical fields and transverse fluctuations we showed in PW12a that only Hall diffusion assists the instability, with ambipolar and Ohm diffusion (which combine as Pedersen diffusion) only able to damp waves. In the present work for a more general field topology and oblique wave vector, we shall see that both ambipolar and Hall diffusion can assist the instability.  Finally, the general stability criterion for a magnetic-diffusion-dominated plasma in the presence of shear flows is presented in this work.

The paper is organised in the following fashion. The basic set of equations and dispersion relation are given in Sec.~2; in subsection 2.1 we give the linearised equations in terms of the diffusion tensor and derive the general dispersion relation. In Sec.~3 the general stability criterion is described and the maximum growth rate of the instability is derived.  The expressions for the maximum growth rate and critical wavelength are given in limiting cases. In Sec.~4, various limiting cases of the dispersion relation are discussed. In Sec.~5 applications of the results are outlined. Finally, in Sec.~6 we give a brief summary of the main results.

\section{Basic model}
The photosphere--chromosphere plasma consists primarily of electrons, protons, singly ionized metallic ions, H, He I, He II, and He III. We shall ignore the distinction between hydrogen and helium and assume that the photosphere and chromosphere are comprised of electrons, singly charged ions and neutral hydrogen.  Although the fundamental set of equations describing a partially ionised plasma was formulated more than 50 years ago \citep{C57,BR65}, we shall adopt the convenient single-fluid formulation given by \cite{P08a} obtained by clearly elucidating the relevant spatial and temporal scales of partially ionised plasmas. 

In the low frequency limit, the collisional dynamics and fractional ionisation can be incorporated in the basic set of equations without having to deal with the complexity of three--fluid equations. Thus the continuity equation of the bulk plasma fluid is 
\bq
\frac{\partial \rho}{\partial t} + \grad\cdot\left(\rho\,\v\right) = 0\,.
\label{eq:cont}
\eq
Here $\rho = \rho_i + \rho_n$ is the bulk mass density and  $\rho_{i\,, n} = m_{i\,,n}\,n_{i\,,n}$ are the ion and neutral mass densities with $m_{i\,,n}\,, n_{i\,,n}$ as the ion and neutral mass and number densities respectively; $\v = (\rho_i\,\vi + \rho_n\,\vn)/\rho$ is the bulk velocity, and, $\vi$ and $\vn$ are bulk velocities of the ion and neutral fluids respectively. The momentum equation is 
\bq
\rho\,\frac{d\v}{dt}=  - \nabla\,P + \frac{\J\cross\B}{c}\,,
\label{eq:meq}
\eq
where $\J = e\,n_e\,\left(\vi - \ve\right)$ is the current density, $\B$ is the magnetic field and $P = P_e + P_i + P_n$ is the total pressure.
The induction equation is
\begin{eqnarray}
\delt \B = \curl\left[
\left(\v\cross\B\right) - \frac{4\,\pi\,\eta_O}{c}\,\J - \frac{4\,\pi\,\eta_H}{c}\,\J\cross\hB
\right. \nonumber\\
\left.
+ \frac{4\,\pi\eta_A}{c}\,
\left(\J\cross\hB\right)\cross\hB
\right]\,,
\label{eq:ind}
\end{eqnarray}
where $\hB = \B /B$ is the unit vector along the magnetic field, 
and the Ohm ($\eta_O$), ambipolar ($\eta_A$) and
Hall ($\eta_H$) diffusivities are 
\bq \eta_O =
\frac{c^2}{4\,\pi\sigma}\,\,, \eta_{A} =
\frac{D^2\,B^2}{4\,\pi\,\rho_i\,\nu_{in}}\,, \eta_H =
\frac{c\,B}{4\,\pi\,e\,n_e}\,.
\label{eq:diffu}
 \eq
Here 
\bq
\sigma = \frac{c\,e\,n_e}{B}\, \left[\frac{\omega_{ce}}{\nu_{e}} + \frac{\omega_{ci}}{\nu_{i}} \right]
\eq 
is the parallel conductivity, $\omega_{cj} = e\,B/m_j\,c$ is the particle$\textquoteright$s cyclotron frequency where $e \,,B\,,m_j\,,c$ denotes the charge, magnetic field, mass and speed of light respectively and  $D = \rho_n/\rho$ is the ratio of neutral and bulk densities. For electrons $\nu_{e} \equiv \nu_{en}$ and for ions $\nu_{i} \equiv \nu_{in}$. Although $\nu_{ee}\,,\nu_{ei}\,,\nu_{ii}\,\mbox{and}\,,\nu_{ie}$ can become comparable to  $\nu_{en}$ (see Table.~1), it is the neutral-plasma collision that gives rise to ambipolar and Hall diffusion in the medium.\footnote{To leading order collisions between like plasma particles $\nu_{ee},\,\nu_{ii}$ do not cause diffusion \citep{LR56}.}  The electron-ion collision contributes to Ohm diffusion. 

Defining the plasma Hall parameter $\beta_j$ as
\bq
\beta_j = \frac{\omega_{cj}}{\nu_{jn}}\,,
\eq
and the Hall frequency
\bq
\omega_H  = \frac{\rho_i}{\rho}\,\omega_{ci} \,, 
\eq  
the diffusivities in Eq.~(\ref{eq:diffu}) can be written in the compact form \citep{P08a}
\bq
\eta_H = \left(\frac{v_A^2}{\omega_H}\right)\,,\quad
\eta_A = D\,\left(\frac{v_A^2}{\nu_{ni}}\right)\,,\quad
\eta_O = \beta_e^{-1}\,\eta_H\,,
\label{eq:ddf}
\eq
where $\nu_{ni} = \rho_i\,\nu_{in} / \rho_n$ and $v_A = B / \sqrt{4\,\pi\,\rho}$.

The induction equation (\ref{eq:ind}) can be written explicitly in terms of the fluid and field velocities as \citep{WS12}
\bq
\delt \B = \curl\left[
\left(\v + \v_B\right)\cross\B - \frac {4\,\pi\,\eta_O}{c}\,\Jpa\right]\,,
\label{eq:ind1}
\eq
where the field drift velocity $\v_B$ is   
\bq
\v_B = \eta_P\,\frac{\left(\grad\cross\B\right)_{\perp}\cross\hB}{B} -√± 
\eta_H\,\frac{\left(\grad\cross\B\right)_{\perp}}{B}\,, 
\label{eq:mdf}
\eq
 and 
\bq
\etaP = \eta_O + \eta_A
\eq 
is the Pedersen diffusivity. The parallel and perpendicular current components in these equations refer to their orientation with respect to the background magnetic field, i.e.
\bq
\Jpa = \left(\J\cdot\hB\right)\,\hB\,,\quad \Jpe = \J - \Jpa\,.
\eq
As we shall see, writing the induction equation in terms of magnetic drift velocity, as in Eq.~(\ref{eq:ind1}), is particularly convenient for the linearization which we will conduct shortly. 

We note from Eq.~(\ref{eq:diffu}) that the magnitude of Ohm diffusion is independent of magnetic field strength, whereas Hall diffusion has a linear dependence on the field, i.e. $\eta_H \propto B$, and the dependence of ambipolar diffusion on the magnetic field is quadratic, i.e. $\eta_A \propto B^2$.  Thus, as the magnetic field strength in a flux tube decreases with increasing altitude, the drop in the Hall diffusivity will not be as severe as for ambipolar diffusion.  The altitude dependence of the diffusivities is easily quantified once the variation of the magnetic field with altitude is specified; for this purpose we adopt a power-law variation of the magnetic field strength with neutral number density $n_n$, i.e.  
\bq
B = B_0\,\left(\frac{n_n}{n_0}\right)^{0.3}\,,
\label{eq:scl}
\eq 
where $n_0$ is the number  density of neutrals at the surface of the photosphere (i.e.~$h = 0$) and $B_0 = 1.2\,\mbox{kG}$ is the typical value of the field at intergranular boundaries. Such a field profile captures the height dependence of the observed field in flux tubes \citep{M97}.  We note that the above scaling of the magnetic field differs somewhat from the mass density scaling \citep{LAK05} although we have retained the same power law index $0.3$.   

The neutral number density and fractional ionisation (Model C, VAL81), the ratio of the neutral and bulk densities $D$, and various collision frequencies are given in Table~1. The metallic ion-neutral collision cross-section is not known, 
so we have employed the Messy-Mohr analytic formula for the cross-section (see appendix). The resulting plasma-neutral collision frequency is an order of magnitude smaller than previously used values by \cite{P08b} where the larger ion-H$_2$ collision cross-section was adopted. 

\begin{figure}
%\vspace*{4cm}
\includegraphics[scale=0.31]{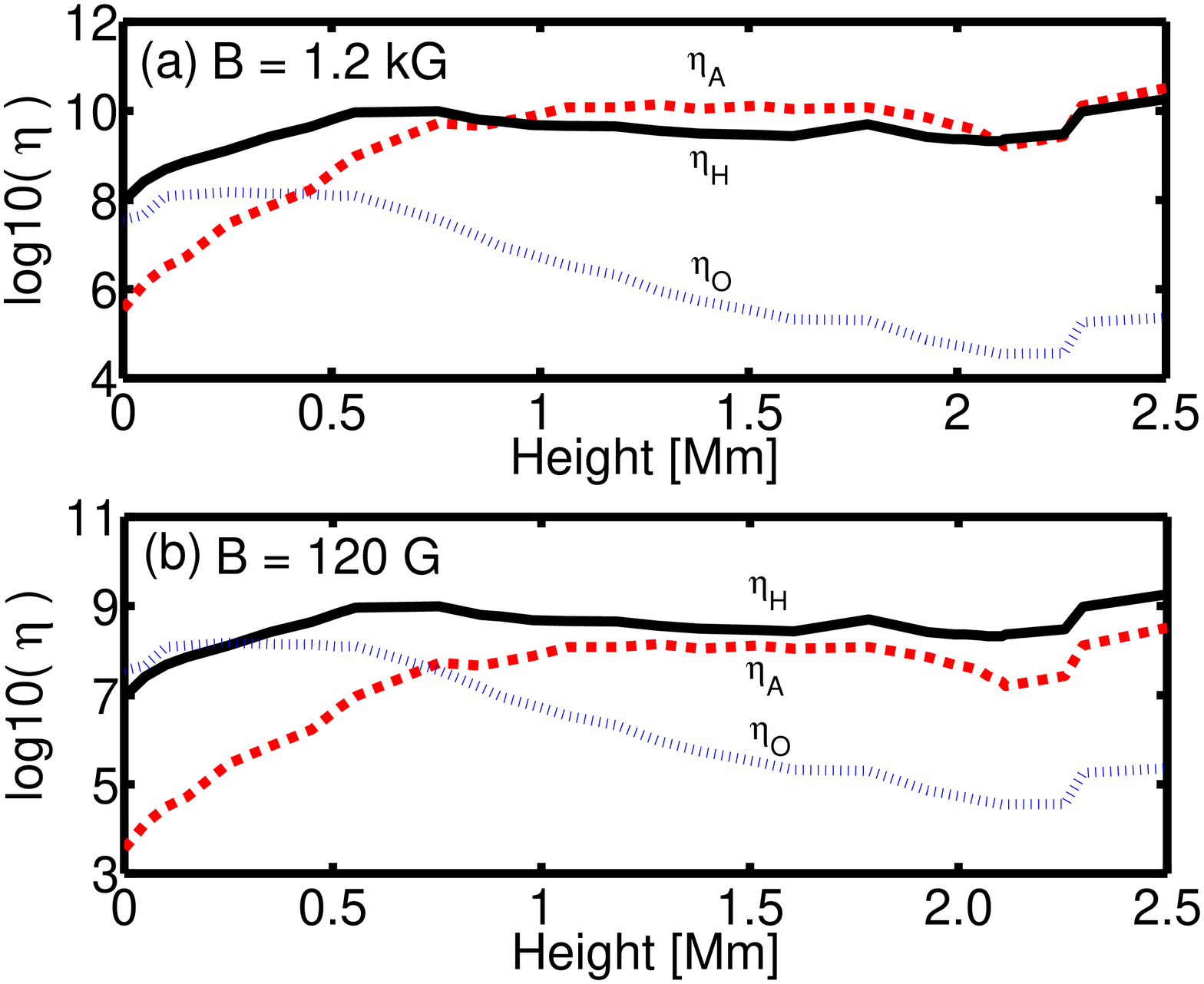}
\caption{The profiles of Ohm ($\eta_O$), Hall ($\eta_H$) and ambipolar ($\eta_A$) diffusivity in the photosphere√±-chromosphere are shown for (a) $B_0 = 1.2\,\mbox{kG}$ and (b) $B_0 = 120 \,\mbox{G}$ (see eq [\ref{eq:scl}]).}
\label{fig:DF}  
\end{figure}
We utilise the above magnetic field profile, along with the collision frequencies given in Table~1, to compute the diffusivities; these are given in Table 2.  In Fig~\ref{fig:DF} we plot the diffusivities against height, demonstrating that Hall diffusion dominates Ohm and ambipolar diffusion in the photosphere and lower chromosphere in the intense field regions (Fig~\ref{fig:DF}a). However, in the weak field regions ($\sim 100\,\mbox{G}$ at $ h = 0$) Ohm and Hall are comparable below $ 0.3\,\mbox{Mm}$, but Hall diffusion still dominates the entire photosphere-chromosphere (Fig~\ref{fig:DF}b).  

It has been suggested in the past that Pedersen diffusion (which is the sum of Ohmic and ambipolar diffusions) could play an important role in chromospheric heating \citep{G04}. We find that in sunspots, active regions, pores and intergranular regions where the field is intense, Hall and ambipolar will dominate Ohm diffusion. Clearly, realistic modelling of energetic processes such as chromospheric heating, CMEs etc. must include these diffusive processes. While Hall diffusion is non--dissipative in nature, it does have a role in extracting energy from the convective motion of the largely neutral medium. As we have shown recently \citep{PW12a}, Hall diffusion in the presence of shear flow  destabilises low frequency fluctuations which may lead to the turbulent cascade of convective energy to smaller scales where dissipation can convert it to heat. Therefore, it is quite plausible that the Hall effect may play an important role in heating the chromospheric plasma.  

\begin{table*}
% \centering
% \begin{minipage}{140mm}
\caption{\label{tab:table1} Height dependence of neutral number density $n_n$, fractional ionisation $X_e \equiv n_e/n_n $ [Model C, \citep{VAL81}], ratio of neutral to bulk mass density $D = \rho_n / \rho$, and the ion-neutral, electron-neutral, and electron-ion collision frequencies. We have assumed $B = B_0 \left( n_n / n_0\right)^{0.3}$ with $B_0 = 1.2\,\mbox{kG}$ and $m_i = 30\,m_p$ and $m_n = 2.3\,m_p$ where $m_p$ is the proton mass.}
  \begin{tabular}{|@{}llrr||rrrrl}
    \toprule[0.12em]
h\,(km) & $n_n(\mbox{cm}^{-3})$  & $X_e$ &$D$ &  $\nu_{in}\,(\mbox{Hz})$ & $\nu_{en}\,(\mbox{Hz})$& $\nu_{ei}(\mbox{Hz})$ \\
\midrule[0.12em]
 $0$ & $1.2\cdot10^{17}$ & $5.5\cdot10^{-3}$ & $1$& $10^{8}$ & $7.4\cdot 10^{9}$ & $1.3\cdot 10^{9}$  \\
$250$ & $2.3\cdot10^{16}$ & $10^{-4}$&$1$& $10^{7}$ & $1.3\cdot 10^{9}$ & $1.2\cdot 10^{8}$   \\
$515$ & $2.1\cdot10^{15}$ & $1.2\cdot10^{-4}$ & $1$ & $10^{6}$ & $10^{9}$& $1.3\cdot 10^{7}$  \\
$1065$ & $1.7\cdot 10^{13}$ & $10^{-2}$ & $0.91$ & $10^{4}$ & $10^{6}$& $3.1\cdot 10^{6}$ \\
$1515$ & $\quad\quad10^{12}$ & $6\cdot10^{-2}$ & $0.53$ & $8\cdot 10^{2}$ & $6.6\,\cdot 10^{4}$&$2.0\,\cdot 10^{6}$  \\
$2050$ & $7.7\cdot 10^{10}$ & $5\cdot10^{-1}$ & $0.12$&$60$ & $5.3\cdot 10^{3}$&$9.6\cdot 10^{5}$  \\
$2298$ & $3.2\cdot 10^{9}$ & $1\cdot10^{0}$ & $0.07$&$6.3$ & $10^{3}$&$10^{5}$  \\
$2543$ & ${}\quad\quad10^{9}$ & $1.2\cdot 10^{0}$ & $0.05$ & $6.3$&$5.3\cdot 10^{2}$ & $3.2\cdot 10^{2}$ \\
\bottomrule[0.12em]
\end{tabular}
%\end{minipage}
\end{table*}
\begin{table*}
 %\centering
 %\begin{minipage}{140mm}
  \caption{\label{tab:table1} Ohm, $\eta_O$, ambipolar, $\eta_A$ and Hall, $\eta_H$ diffusivities at different altitudes, based on the collision frequencies in Table~1.}
  \begin{tabular}{llll@{\qquad\qquad}llll}
    \toprule[0.12em]
h\,(km) & $\eta_O(\mbox{cm}^{2} / \mbox{s})$  & $\eta_A(\mbox{cm}^{2} / \mbox{s}) $ &$\eta_H(\mbox{cm}^{2} / \mbox{s})$ &  $ h\,(km)$ & $\eta_O(\mbox{cm}^{2} / \mbox{s})$& $\eta_A(\mbox{cm}^{2} / \mbox{s})$& $\eta_H(\mbox{cm}^{2} / \mbox{s})$\\
\midrule[0.12em]
 $0$ & $3.6\cdot10^{7}$ & $3.6\cdot10^{5}$ & $9.3\cdot10^{7}$& $1180$ & $2.1\cdot 10^{6}$ & $1.2\cdot 10^{10}$ &$4.5\cdot 10^{9}$  \\
$100$ & $1.2\cdot10^{8}$ & $3.1\cdot10^{6}$&$4.8\cdot10^{8}$& $1380$ & $5.5\cdot 10^{5}$ & $1.1\cdot 10^{10}$ &$3.1\cdot 10^{9}$ \\
$250$ & $1.5\cdot10^{8}$ & $2.9\cdot10^{7}$ & $1.3\cdot10^{9}$ & $1605$ & $2.1\cdot10^{5}$& $1.1\cdot 10^{10}$ &$2.7\cdot 10^{9}$  \\
$515$ & $1.2\cdot 10^{8}$ & $5.2\cdot10^{8}$ & $7.1\cdot10^{9}$ & $1925$ & $7.3\cdot10^{4}$& $7.2\cdot 10^{9}$ &$2.6\cdot 10^{9}$   \\
$555$ & $1.2\cdot10^{8}$ & $9.4\cdot10^{8}$ & $9.2\cdot10^{9}$ & $2016$ & $5.1\,\cdot 10^{4}$&$4.4\,\cdot 10^{9}$ & $2.3\cdot 10^{9}$  \\
$755$ & $3.6\cdot 10^{7}$ & $5.3\cdot10^{9}$ & $9.8\cdot10^{9}$&$2104$ & $3.5\cdot 10^{4}$&$2\cdot 10^{9}$ & $2.1\cdot 10^{9}$  \\
$855$ & $1.5\cdot 10^{7}$ & $4.6\cdot10^{9}$ & $6.3\cdot10^{9}$&$2255$ & $3.6\cdot 10^{4}$&$2.7\cdot 10^{9}$ & $3\cdot 10^{9}$   \\
$980$ & $5.9\cdot10^{6}$ & $7.6\cdot 10^{9}$ & $4.8\cdot10^{9}$ & $2543$&$2.4\cdot 10^{5}$ & $4\cdot 10^{10}$ &$2.1\cdot 10^{10}$ \\
\bottomrule[0.12em]
\end{tabular}
\label{table:dift}
%\end{minipage}
\end{table*}

How much diffusion is too much in the solar atmosphere? For example, if magnetic diffusion dominates fluid convection in the induction equation~(\ref{eq:ind}), the magnetic field will be poorly coupled to the plasma \citep{W07}. Thus, we shall compare the advection term $\curl\left(\v\cross\B\right) \sim v\,B / L$ with the diffusion terms in the induction Eq.~(\ref{eq:ind}) by defining magnetic Reynolds numbers
\bq
\mbox{Rm} = \frac{v\,L}{\eta_O}\,,\mbox{Am} = \frac{v\,L}{\eta_A}\,,\textrm{and}\, \mbox{Hm} = \frac{v\,L}{\eta_H}\,.
\eq
Note that both Am and Hm depend on how well the plasma is coupled to the magnetic field since both ambipolar and Hall diffusion depend on the magnetic field and $\eta_A = \beta_i\,\eta_H$ and $\eta_H = \beta_e\,\eta_O$. Thus
\bq
\mbox{Am} = \mbox{Rm} / \left(\beta_i\,\beta_e\right)\,,\quad\mbox{Hm} = \mbox{Rm} / \beta_e\,.
\eq  
The dependence of Am and Hm on the plasma Hall parameters $\beta_j$ is not surprising given that both ambipolar and Hall diffusion arise from the magnetisation of the medium. This also explains the inherently different nature of Ohm and ambipolar diffusion: whereas Ohm diffusion acts isotropically, ambipolar diffusion owing to its dependence on the magnetic field is anisotropic. As we shall see, in the presence of shear flow, the anisotropic nature of ambipolar diffusion is at the centre of wave destabilization.     

The height dependence of Rm, Hm and Am for both kG and 0.1 kG fields were discussed in PW12a. It was shown that when $v \sim v_A$ (where $v_A$ is the \alf speed), $\mbox{Rm} \gg 1$ suggesting that the Ohm diffusion is unimportant in comparison with the advection term. In contrast, Hm and Am are three to four orders of magnitude smaller than Rm. The recent 2D numerical simulation of the partially ionized solar atmosphere also suggests that in the weak field  regions ($\lesssim 100$ G) in the chromosphere, Hall diffusion is two orders of magnitude larger than Ohm diffusion  whereas ambipolar diffusion is four to six order of magnitude larger than Ohm diffusion \citep{SK12}.  Clearly, the ambipolar and Hall diffusion terms are of critical importance in the induction equation.

\subsection{Dispersion relation}
Magnetic elements in the photosphere-chromosphere are highly dynamic, being accompanied by numerous flows with different spatial and temporal scales. Recent numerical simulations of umbral magnetoconvection \citep{C12} suggest that the dynamical scale over which Hall diffusion can generate magnetic and velocity fields are much smaller and faster than the spatial and temporal scale of a typical flux tube ($\sim$ 10--20 km and $\sim 300\,\mbox{s}$   cf. $\gtrsim$ few hundred km and $\sim$ few days, respectively). The simulation results are easily scalable to $2\,\mbox{km}$ with temporal scale $\sim 2$\,\mbox{s}. We note that at present the best achievable resolution is $90\,\mbox{km}$ \citep{B08}.  

As the spatial scale over which the flow and field generation occurs is much smaller than the typical diameter of a flux tube,  we shall approximate part of the cylindrical tube by a planar sheet and work in  Cartesian coordinates where $x\,,y\,,z$  correspond to the local  radial, azimuthal and vertical directions.  We assume an initial homogeneous state with azimuthal shear flow $\v = {v_0}^{\prime}\,x\,\bmath{y}$. The magnetic field in the intergranular lanes at the network boundaries is clumped into elements or flux tubes that are generally vertical \citep{M97, H09} but highly inclined fields have also been reported in the literature \citep{S87a, S87b}.  The internetwork magnetic elements have predominantly horizontal field \citep{H09, S12}. Therefore, we shall assume a uniform background field that have both azimuthal as well as a vertical component, i.e. $\B = (0, B_y, B_z)$.

The focus of the present investigation is the low frequency behaviour of the medium, and thus, we shall work in the Boussinesq approximation limit which is valid if the motion in the medium is very slow \citep{SV60}. Thus linearising Eqs.~(\ref{eq:cont}) and (\ref{eq:meq}) and assuming an axisymmetric perturbations of the form $\exp \left(i \,\k  \cdot x + \sigma \, t\right)$, with $k = (k_x , 0 , k_z)$, in the Boussinesq approximation, we get
\bq
\k\cdot\dv = 0\,.
\label{eq:cd}
\eq
\begin{eqnarray}
\sigma \,\dvx = - \,i\,k_x\,\frac{\delta p}{\rho} + \frac{i}{4\,\pi\,\rho} 
\left[\left(\k\cdot\B\right)\,\dBx - \left(\B\cdot\dB\right)\,k_x\right]\,, \nonumber\\
\sigma\,\dvy + \vp\, \dvx = \frac{i}{4\,\pi\,\rho} \left(\k\cdot\B\right)\,\dBy\,,\nonumber\\
\sigma\,\dvz = - i\,k_z\,\frac{\delta p}{\rho} + \frac{i}{4\,\pi\,\rho} \left[ \left(\k\cdot\B\right)\,\dBz - \left(\B\cdot\dB\right)k_z\right]\,.
\label{eq:cfm}
\end{eqnarray}

%\begin{eqnarray}
%\end{eqnarray}

Eliminating the pressure perturbation in favour of velocity and making use of Eq.~(\ref{eq:cd}), from the preceding equation we get for the $(x\,,y)$ components 
\bq
\hdv
 = \frac{i\,k\,v_A\,\mu}{\sigma^2}\,
\left(
\begin{array}{cc} \sigma   &  0\\
                   - \vp & \sigma
  \end{array}
\right)\,\hdB\,.
\label{eq:lin1}
\label{eq:Lm}
\eq
Here $\hdv = \dv / v_A\,$ and $\hdB = \dB / B\,$, $\vp = d\,v(x) / dx$, $\mu = \left(\hk \cdot \hB\right) \equiv \kz\,\Bz$,  $\hk = \k / k$ and $\kz = k_z /k$. 
Since 
\begin{eqnarray}
\dvB = i\,k\,\left[\etaH \left\{\mu\,\hdB\cross\hB - \left(\hB\cdot\hdB\right)\hk\times\hB\right\} \right.\nonumber\\
\left.
+ \etaP\left\{\mu\,\dB - \left(\hB\cdot\hdB\right)\hk\right\} 
\right] \,,
\label{eq:lmdf}
\end{eqnarray}
the linearised induction equation ($x\,,y$ components) can be written as
\begin{eqnarray}
\left[
\left(
\begin{array}{cc} \sigma   &  0\\
                  - \vp & \sigma
  \end{array}
\right) 
  + \frac{k^2\,v_A^2\,\mu^2}{\sigma^2}\,\left(
\begin{array}{cc} \sigma   &  0\\
                  - \vp & \sigma
  \end{array}
\right) \right.\,,\nonumber\\
\left. + k^2\,
\left(
\begin{array}{cc} \eta_{xx}   & \eta_{xy}\\
                   \eta_{yx} & \eta_{yy}
  \end{array}
\right)\,\right]\hdB = 0
\,,
\label{eq:lin2}
\end{eqnarray}
where $\bmath{\eta}$ is the diffusivity tensor with following components 
\begin{eqnarray}
\eta_{x\,x} = \eta_O + \Bz^2\,\eta_A\,,\quad \eta_{x\,y} = s\,\eta_H + g\,\eta_A\,,\nonumber\\
\eta_{y\,x} = \left(g\,\eta_A √±- s\,\eta_H\right) / \kz^2\,,  
\quad
\eta_{y\,y} = \eta_O + \left(1 -\kx^2\,\Bz^2\right)\,\eta_A\,.
\label{eq:amt}
\end{eqnarray}
Here 
\bq
g = - \kx\,\kz\,\By\,\Bz\,,
\eq 
and the helicity is $s = \mu\,\kz \equiv \Bz\,\kz^2$. We note that for a purely vertical field the sign of helicity $s$ is determined by the projection of the vertical magnetic field on the vorticity, $\curl\v_0$.   

The following dispersion relation can be derived from (\ref{eq:lin2})
\bq
\hs^4 + \hk^2\,\left(\hP + \hT\right)\,\hs^3 + C_2\,\hs^2 + C_1\,\hs + C_0 =0\,,
\label{eq:mdr}
\eq 
where 
\begin{eqnarray}
C_2  & = & \hk^4\,\left(\hP\,\hT + \mu^2\,\hH^2\right)  +  \hk^2\,\left[2\,\mu^2 - \alpha\, \left(g\,\hA + s\,\hH \right)\right]\,, 
\nonumber\\
C_1 & = & \hk^4\,\mu^2\,\left(\hP + \hT\right)\,, \nonumber\\
C_0  & = & \hk^4\,\mu^2\,\left[\mu^2 - \alpha\,\left(g\,\hA + s\,\hH \right)\right]\,,
\end{eqnarray}
$\alpha = - \vp / \ |\vp| \equiv \pm 1 $ and 
\bq
\eta_T = \eta_O + \mu^2\,\eta_A\,.
\label{eq:dpt}
\eq
We have used the following normalisation in the above equations 
\begin{eqnarray}
\hs = \frac{\sigma}{\avp}\,,\quad 
\hk = \frac{k\,v_A}{\avp}\,,\quad 
\het = \frac{\eta\,\avp}{v_A^2}\,.
\label{eq:scll}
\end{eqnarray} 
Eq.~(\ref{eq:mdr}) reduces to the known dispersion relation \citep{K08} when $D = 1$ and $\eta_O = 0$. When $D = 0$, ambipolar diffusion drops out of the dispersion relation, as expected.

\section{Non-ideal MHD instabilities}
A general stability criterion in the magnetic diffusion dominated plasma can be derived from the dispersion relation, Eq.~(\ref{eq:mdr}) by recasting it in the following simple form
\bq
a\,\hk^4 + b\,\hk^2 + c = 0\,,
\label{eq:dnc}
\eq
where the coefficients $a$, $b$ and $c$ are
\bq
a = \beta - \gamma\,\mu^2\,,\quad b = \delta - \gamma\,\hs^2\,, 
\quad c = \hs^4\,,
\label{eq:coeff1}
\eq
with 
\begin{eqnarray}
\beta &=& \left(\mu^2\,\hH^2 + \hP\,\hT\right)\,\hs^2 + \left(\hP + \hT\right)\,\mu^2\,\hs + \mu^4\,, 
\nonumber\\
\delta &=& \left[\left(\hP + \hT\right)\,\hs + 2\,\mu^2 \right]\,\hs^2\,, 
\nonumber\\
\gamma &=& \alpha\,\left(g\,\hA + s\,\hH\right)\,.
\label{eq:coeff2} 
\end{eqnarray}
The dispersion relation, Eq.~(\ref{eq:dnc}) can also be recast as
\bq
\beta\,\hk^4 + \delta\,\hk^2 + \hs^4 = 
\gamma\,\hk^2\,\left[\hs^2 + \mu^2\,\hk^2\right]\,.
\label{eq:dnc1}
\eq
Thus in the neighbourhood of $\hs \sim \hk \ll 1 $ where 
\bq
\beta \sim \mu^4\,,\quad \delta \sim 2\,\mu^2\,\hs^2\,,
\eq
Eq.~(\ref{eq:dnc1}) becomes
\bq
\mu^2\,\hk^2 + \hs^2 = \gamma^2\,\hs^2\,,
\eq 
which can be written as 
\bq
\gamma - \mu^2 = \left(\frac{\hs}{\hk}\right)^2\,.
\label{eq:sfc}
\eq
From Eq.~(\ref{eq:sfc}) we see that for positive $\sigma$, the right hand side is positive. This implies that $\gamma - \mu^2 > 0$. Therefore, we arrive at the general stability criterion which states that if 
\bq
\alpha\,\left(g\,\hA + s\, \hH \right) > \mu^2\,,
\label{eq:pwd}
\eq
the waves are unstable in the medium. Ohm diffusion does not appear in the above expression, which is not surprising considering that the above criterion pertains to long wavelength fluctuations. For definiteness, in the subsequent analysis, we shall assume $\alpha = 1$. 

The diffusion-shear instability is caused by a competition between the fluid advection and field drift in the plasma. This can be seen from the $y$ and $x$ components of Eqs.~(\ref{eq:lin1}) and (\ref{eq:lmdf}) respectively, which suggest that when $\sigma = 0$, the advection of fluid in the $x$ direction, $\hdvx$ is equal and opposite to  the magnetic field drift velocity $\hdv_{Bx}$. By combining the $y$ component of Eqs.~(\ref{eq:lin1}) and the $x$ component of Eq.~(\ref{eq:lmdf}) we get 
\bq
\hdvx + \hdv_{Bx} = i\,\hk\,\mu\,\left[
\frac{-\gamma + \mu^2}{\alpha\,\mu^2}\,\hdB_y + \frac{1}{\kz^2}\,\hA\,\hdB_x 
\right]\,.
\label{eq:rbal}
\eq
Here we have neglected Ohm diffusion. By setting $- \gamma + \mu^2 = 0$ near the marginal state, when $\hA \ll 1$, we get $\hdvx + \hdv_{Bx} \approx 0$. This provides a simple physical explanation of how magnetic diffusion helps the shear flow to destabilise the waves. The outward slippage of the field in the {x}--direction due to magnetic diffusion weakens the magnetic tension force. As a result the magnetic restoring force in the wave is dominated by the inertial force, resulting in an increased inward drift of the fluid elements. This is how magnetic diffusion assists the waves to grow in the presence of a shear flow. 

The stability criterion, Eq.~(\ref{eq:pwd}) in the dimensional form becomes
\bq
- \vp\,\left(g\,\eta_A + s\,\eta_H\right) > \mu^2\,\v_A^2\,,
\label{eq:pwd1}
\eq
which suggests that when $- \vp > 0$, the linear combination of ambipolar and Hall diffusivities multiplied by suitable topological factors $g$ and $s$ must exceed the square of the oblique \alf speed.  The above equation provides a simple stability criterion of diffusive medium. For example, low frequency fluctuations in the Hall--Ohm dominated photosphere-lower chromosphere ($\eta_A = 0$) are unstable if 
\bq
- \vp\eta_H > \Bz\,v_A^2\,, 
\eq
or, in terms of Hall frequencies  
\bq
- \vp > \Bz\,\omega_H \equiv \omega_{HZ}\,.
\label{eq:CH1}
\eq
Here we have used $\eta_H = v_A^2 / \omega_{H}$ (PW12a) and $\Bz\,\omega_{H} =\omega_{HZ}$ is the Hall frequency defined in terms of the vertical field.   
Clearly if the Hall frequency is less than the shear frequency $- \vp$, Hall diffusion can drive the shear flow instability.

When $\vp\,\eta_H / v_A^2 = - \Bz$, since $\eta_A > 0$, ambipolar diffusion can drive shear flow instability provided $g > 0$. For purely vertical fields or, transverse fluctuations (vertical wavevector) when $g = 0$, ambipolar diffusion can only cause damping of waves. Therefore, the question of ambipolar diffusion assisting the shear flow instability is inherently linked to the ambient field geometry and wave obliqueness which together is encapsulated in the topological factor $g$. The important role of $g$ in the ambipolar diffusion driven shear flow instability was discovered by \cite{D04}. 

How does ambipolar diffusion help to drive the shear flow instability? In order to see this, we first note that when the wavevector is perfectly aligned to the ambient magnetic field, i.e. $\mu \equiv \kz\,\Bz = 1$, $\eta_T \equiv \eta_P$ in Eqs.~(\ref{eq:mdr})--(\ref{eq:dpt}), Ohm and ambipolar diffusion combines together as Pedersen diffusion and their effect on the wave propagation is identical--they both cause wave damping. Only when $\mu \neq 1$, this diffusive degeneracy is lifted and Ohm and ambipolar diffusion can no longer be combined together as single diffusion. After removal of the degeneracy whereas Ohm diffusion still causes isotropic damping of the waves, damping by the ambipolar diffusion becomes anisotropic. Therefore, the non--vertical magnetic field and oblique wavevector (which is crucial in removing this degeneracy) plays an important role in the ambipolar diffusion driven shear instability. This can also be seen from the genelarlized induction Eq.~(\ref{eq:ind}) if we rewrite the electric field as
\begin{eqnarray}
\E^\prime = \frac{4\,\pi}{c^2}\left\{\eta_O + \left[ 1 - \left(\bmath{j}\cdot\hB \right)^2 \right]\,\eta_A\,\right\}\,\J + \frac{4\,\pi}{c^2}\,\eta_H\,\left(\J\cross\hB \right)\nonumber\\
+ \frac{4\,\pi}{c^2}\,\eta_A\, \left(\bmath{j}\cdot\hB \right)\,\left[\bmath{j}\cross\left(\J\cross\hB \right)\right]\,,
\label{eq:ef}
\end{eqnarray}   
where the electric field is written in the neutral frame, $\bmath{j} = \J / |J|$. It is clear from the preceding equation that when the medium is threaded only by the vertical field and wave is propagating along the field, i.e. $\bmath{k} = (0, 0, k_z)$, $\bmath{j} = (j_x, j_{y}, 0)$, both Ohm and ambipolar diffusion cause damping of the waves since $\left(\bmath{j}\cdot\hB \right) = 0$ and  $\E^\prime\cdot\J = \eta_P\,\J^2$. However, when $\mu \neq 1$, the last term in Eq.~(\ref{eq:ef}) can help fluctuations to grow since 
\begin{eqnarray}
\E^\prime \cdot \J = \frac{4\,\pi}{c^2}\left\{\eta_O + \left[ 1 - \left(\bmath{j}\cdot\hB \right)^2 \right]\,\eta_A\,\right\}\,\J^2 \nonumber\\
+ \frac{4\,\pi}{c^2}\,\eta_A\, \left(\bmath{j}\cdot\hB \right)\,\left[\J\cdot \left(\bmath{j}\cross\left(\J\cross\hB \right)\right)\right]\,.
\label{eq:ef1}
\end{eqnarray}
Therefore, ambipolar diffusion plays dual role in a partially ionized medium: whereas, in one direction it can cause dissipation like Ohm, in the other direction the dissipation is considerably smaller. The directional dissipation is the hall mark of ambipolar diffusion which causes it to assist the instability \citep{D04}.  

From the stability criterion Eq.~(\ref{eq:pwd1}) it is clear that the Hall--Ohm stable medium ($- \vp\, \eta_H \leq \Bz\,v_A^2$) may or may not be ambipolar unstable whereas the Hall--Ohm unstable medium ($- \vp\,\eta_H > \Bz\,v_A^2$) can always become ambipolar unstable for non--vertical fields and oblique wavevectors since $g > 0$ can be easily satisfied. The ambipolar diffusion not only drives the shear flow instability when $g > 0$, but also enlarges the parameter space over which Hall can destabilise the waves. Therefore Hall diffusion can drive the shear flow instability when  
\bq
\eta_H > \frac{\Bz\,v_A^2}{- \vp} √±- \frac{g}{s}\,\eta_A\,.
\label{pwd2}
\eq   
Since $\Bz\,,\By \in [0,1]$ and maximum $g = 0.25$, the above inequality implies that for non-zero $\eta_A$, Hall diffusion drives shear instability at much larger negative value than when the field has only a vertical component. 

The stability criterion can also be recast as  
\bq
- \left(\frac{\kx}{\kz}\,\frac{\By}{\Bz}\right) > \frac{1}{\eta_A}\,\left( \frac{v_A^2}{- \vp} - \frac{1}{\Bz}\,\eta_H\right)\,.
\label{eq:pwd2}
\eq       
Assuming a positive left hand side in the preceding equation we see that the above criterion is easily satisfied for non-zero $\eta_A$ when $\kz \rightarrow 0$. Therefore, waves propagating almost along $x$ direction i.e. when the fluctuations are almost magnetosonic are always unstable.
 
The maximum growth rate of the instability can be found by setting the discriminant $b^2 √±- 4\,a\,c = 0$ in Eq.~(\ref{eq:dnc}). This yield
\bq
\hs_0 = \frac{g\,\eta_A + s\,\eta_H}{\left[\left(\eta_P + \eta_T\right) + 2\,\sqrt{\mu^2\,\eta_H^2 + \eta_P\,\eta_T}\right]}\,.
\label{eq:Gmax}
\eq
We see that the maximum growth rate depends on both $g\,\eta_A$ and $s\,\eta_H$ and for comparable $\eta_A$ and $\eta_H$ [as is the case in large part of the solar atmosphere; see Fig.~(\ref{fig:DF})], the bigger contribution to the growth rate comes from the Hall diffusion since the maximum value of parameter $g$  is $0.25$ whereas maximum value of the helicity $s$ is one.

In the absence of Hall and Ohm, the maximum growth rate, Eq.~(\ref{eq:Gmax}) becomes 
\bq
\sigma_0 = \frac{g\,|\vp|}{\left(1 + \mu\right)^2}\,,
\label{eq:amax}
\eq
which suggests that when the field drift is solely due to the ambipolar diffusion the maximum growth rate is independent of the diffusivity  as well as the strength of the background magnetic field. However, the signature of the magnetic field in Eq.~(\ref{eq:amax}) appears through the topological factors $g$ and $\mu$. 

When the field drift in the plasma is solely due to the Hall diffusion, the maximum growth rate Eq.~(\ref{eq:Gmax}) becomes
\bq
\sigma_0 = \frac{|\vp|}{2}\,\kz\,,
\label{eq:hmax1}
\eq
which implies that the waves with $\kz = 1$ are most unstable. A comparison with Eq.~(\ref{eq:amax}) shows that in the purely ambipolar or, the purely Hall case, growth rate is independent of the ambient diffusivity. 

\begin{figure}
%\vspace*{4cm}
 \includegraphics[scale=0.31]{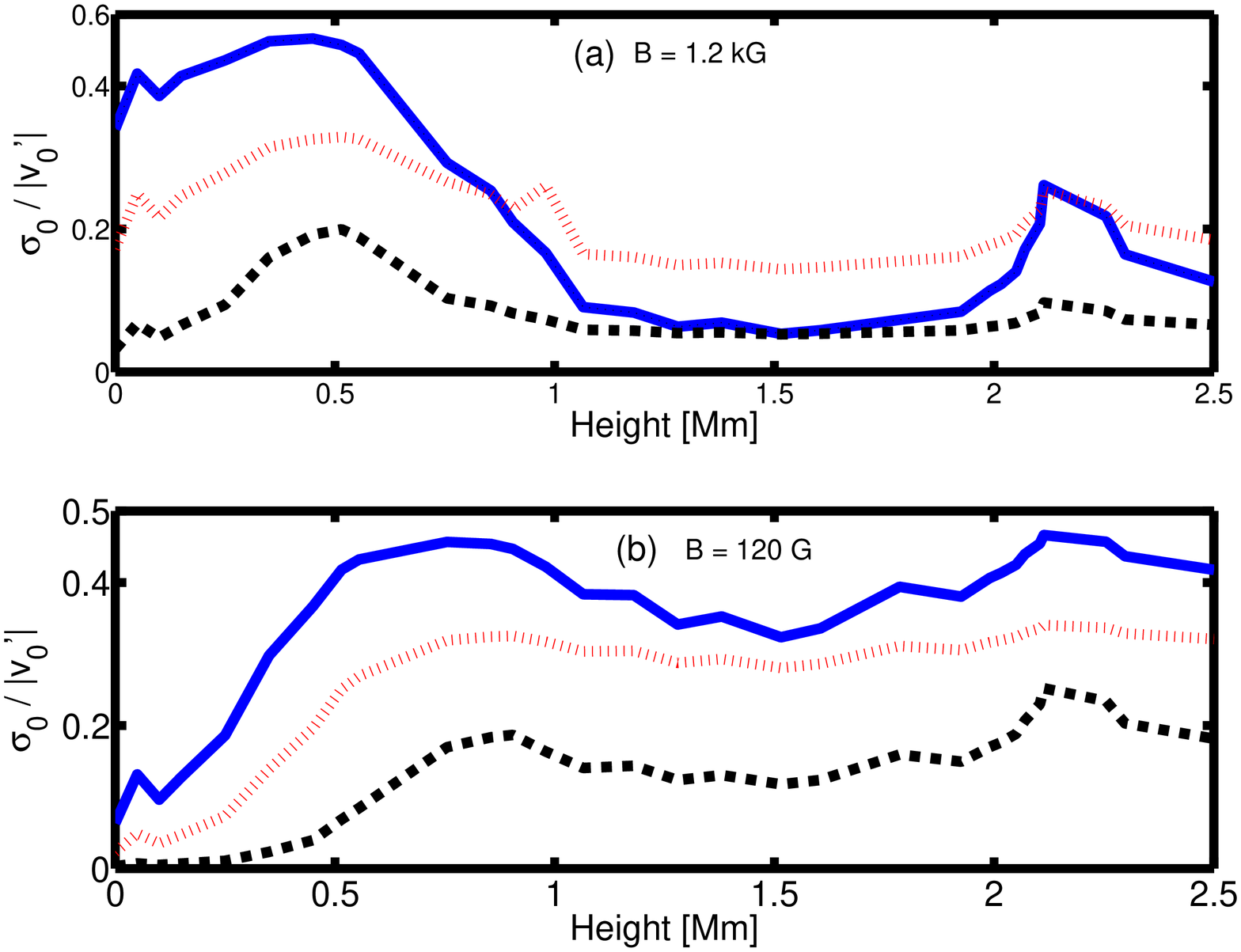}
\caption{The maximum growth rate, Eq.~(\ref{eq:Gmax}) is shown for $\kz = \Bz = 1$ (bold line), $\kz = \kx = \Bz = - \By = 1 / \sqrt{2}$ (dotted line) and $\kz = 1 / \sqrt{2}$ and $\Bz =0.1$ (dashed line) for (a) $\mbox{B} = 1.2\,\mbox{kG}$ and (b) $\mbox{B} = 120 \,\mbox{G}$.}
 \label{fig:DF2}  
\end{figure}
We see from Fig.~\ref{fig:DF2}(a) that for a purely vertical field and a vertical wavevector ($\kz = \Bz = 1$, $g = 0$, bold lines) the instability grows at a maximum rate of $\avp /2$ in the photosphere-lower chromosphere in strong field region. Recall that in this interval Hall is the dominant diffusion mechanism in the network and internetwork regions (Fig.~\ref{fig:DF}) and dissipation due to the Ohm and ambipolar diffusion (which can be combined together as Pedersen diffusion for this topology) is small.

When the field is weak [Fig.~\ref{fig:DF2}(a)] the instability grows close to the maximum rate in the entire photosphere-chromosphere except very close to the surface ($\lesssim 0.2\,\mbox{Mm}$). When $\kz = \kz = \Bz = - \By = 1 / \sqrt{2}$ [dotted curves in Fig.~2(a)--(b)] the growth rate is smaller than the previous case. This is because in this case the instability is not only due to Hall diffusion but is also assisted by ambipolar diffusion through directional dissipation of waves. Therefore, the growth rate is always smaller when both Hall and ambipolar diffusion are present in the medium. With the decreasing vertical field [$\Bz = 0.1$, dash-dot curves in Fig.~2(a)--(b)] the maximum growth rate decreases further owing to smaller $g$ and $s$.   

\begin{figure}
%\vspace*{4cm}
 \includegraphics[scale=0.31]{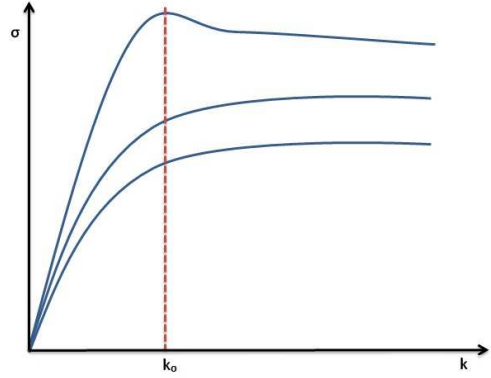}
\caption{Above cartoon depicts growth rate against wavenumber.}
 \label{fig:F3}  
\end{figure}
The most unstable wavenumber corresponding to maximum growth rate Eq.~(\ref{eq:Gmax}) is $\hk_0^2 = - 2\,c / b$ which can be written as
\bq
\hk_0^2 = \frac{\hs_0^2}{\left(\sqrt{\mu^2\,\hH^2 + \hP\,\hT}\right)\,\hs_0 - \mu^2}\,.  
\label{eq:k0s}
\eq
It is clear from preceding equation that when
\bq
g\,\hA + s\,\hH > 2 + \frac{\hP + \hT}{\sqrt{\mu^2\,\hH^2 + \hP\,\hT}}\,, 
\eq
the maximum growth rate occurs at finite $k_0$ [see Fig.~(\ref{fig:F3})]. In the opposite limit $k_0$ becomes imaginary implying that the maximum growth rate occurs at infinity. For typical parameters of network-internetwork magnetic elements, with shear gradient $\vp \sim 10^{-2} \persec$ and $v_A \sim 10^{5}\, \kms$, $k_0$ is imaginary [corresponding to two similar curves in Fig.~(\ref{fig:F3})]. 

\section{Limiting cases of the dispersion relation}
The magnetic field drift in the partially ionized plasma opens up new pathways through which the free shear energy of the fluid can be transferred to the waves (PW12a). To see this, we first note that since $\hs \sim 1$, three terms in Eq.~(\ref{eq:lin2}) are $\sim 1\,,\hk^2\,,\hk^2\,\eta$ and thus, one of the following three scenarios may prevail in a diffusive medium.

{\bf A. {\it Ideal MHD:}}  In this limit $\hk^2 \sim 1 \gg \hk^2\,\het$, i.e. $\het \ll 1$ and last term in Eq.~(\ref{eq:lin2}) can be neglected. However, as has been shown in PW12a, this limit is applicable to the long wavelength fluctuations $\gtrsim 10^3\,\mbox{km}$ and therefore, is not relevant to the present analysis. 

{\bf B. {\it Cyclotron limit:}} In this case $\hk^2\,\het \sim \hk^2 \gg 1$. This is the low frequency limit and first term in Eq.~(\ref{eq:lin2}) can be neglected. The low frequency, a short wavelength ion-cyclotron wave with frequency $\omega_{C} = \mu\, \omega_H$ is the normal mode of the system \citep{PW06, P08a}. 

Assuming $\avp \sim 0.01 \persec$ and $v_A \sim 5\times 10^5\,\kms$ we get 
$\het \equiv \eta\,|\vp| / v_A^2 \sim 10^{-2}-10^{-4} \ll 1$ and thus it was inferred (PW12a) that the cyclotron limit is not valid in the photosphere--chromosphere. However, at an increased shear frequency $\avp \lesssim 1 \persec$ we get $\het \equiv \eta\,|\vp| / v_A^2 \lesssim 1$ and this limit becomes important. Since recent numerical simulation with $10-20\,\mbox{km}$ resolution \citep{C12} can be easily scaled to $2\,\mbox{km}$ which for typical $v_0 \sim 2 \kms$ gives $\avp \sim 1 \persec$, we conclude that the proper analysis of the cyclotron limit will provide important insight to the ongoing numerical simulations of  the photosphere--chromosphere.   
  
Taking $k \rightarrow \infty$, and thus setting $a(\sigma) = 0$ in Eq.~(\ref{eq:dnc}) or, neglecting the first matrix in Eq.~(\ref{eq:lin2}), we get following dispersion relation
\bq
\left(\mu^2\,\hH^2 + \hP\,\hT\right)\,\hs^2 + \mu^2\,\left(\hP + \hT\right)\,\hs = \mu^2\,\left(\gamma - \mu^2\right)\,.
\label{eq:EsL} 
\eq
Positive $\sigma$ requires $\gamma > \mu^2$ since coefficients on the left hand side of preceding equation are positive. Thus the stability criterion in the cyclotron limit is identical to general stability criterion, Eq.~(\ref{eq:pwd}). This is not surprising since above dispersion relation is the short wavelength limit of the general case.  

The growth rate of the instability becomes
\bq
\hs = \frac{\mu^2}{2\,\left(\hP\,\hT + \mu^2\,\hH^2\right)} \left[ - \left(\hP + \hT\right) + \sqrt{\Delta}\right]\,,
\label{eq:gcx}
\eq 
where
\bq
\Delta = \left(\hP - \hT\right)^2 + 4\,\hH^2\left(\gamma √±- \mu^2\right) + 4\left(\frac{\gamma\,\hP\,\hT}{\mu^2}\right)\,.
\label{eq:gcc}
\eq
The above equation acquires a particularly simple form in the purely Hall or ambipolar limits. For example in the Hall diffusion dominated regime, from Eq.~(\ref{eq:gcx}) we get 
\bq
\hs = \left(\frac{\alpha\,s}{\hH} - \frac{\mu^2}{\hH^2}\right)^{1/2}\,,
\eq
which in the dimensional form becomes 
\bq
\sigma = \left[ - \vp\,\frac{s}{\eta_H} - \mu^2\,\frac{v_A^2}{\eta_H^2}
\right]^{1/2}\, v_A\,.
\eq
In the $\mu < 1$ limit above equation can be written as
\bq 
\sigma \approx \left[ - \vp\,\,s\,\omega_H\right]^{1/2}\,.
\label{eq:ic1}
\eq
Thus the growth rate of the ion-cyclotron wave approximately equals the geometric mean of the shear and Hall frequencies and attains a maximum value only for positive helicity $s = 1$.   

In the purely ambipolar diffusion dominated case setting $\eta_O = \eta_H = 0$, we get
\bq
\sigma = \frac{v_A^2}{2\,\eta_A}\,\left[ - \left(1 + \mu^2\right) + 
\sqrt{\left(1-√±\mu^2\right)^2 - \frac{4\,g\,\vp\,\eta_A}{v_A^2}}
\right]\,.
\label{eq:ahfmax1}
\eq
We see from Eq.~(\ref{eq:ahfmax1}) that when $g = 0$, ambipolar diffusion causes only damping. As noted in the previous section, it is only when the topological factor $g$ is non-zero that anisotropic ambipolar diffusion can drive shear flow instability. 

{\bf C. {\it Highly diffusive limit :}} In this limit $\hk^2\,\het \sim 1 \gg \hk^2$, i.e. $\het \gg 1$ and $\hk^2 \ll 1$. Only the first and last term in Eq.~(\ref{eq:lin2}) are retained. For typical values of $\eta$ [Table~(\ref{table:dift})], this limit gives $\lambda \lesssim 6\,\mbox{km}$ which fits within a pressure scale height. Therefore, as noted in our previous work (PW12a), a highly diffusive limit is applicable to the photosphere-chromosphere. 

The dispersion relation in this limit becomes
\bq
\left(\hP + \hT \right)\,\hs\,\hk^2 + \hs^2 = \hk^2\,\left[\gamma -√± \hk^2\,\left(\hP\,\hT + \mu^2\,\hH^2\right)\right]\,,  
\label{eq:HDL1}
\eq
from where we see that for a positive $\sigma$ the left hand side is positive. Thus the right hand side must be positive. Thus we arrive at the following general stability criterion 
\bq
- \vp\left(g\,\eta_A + s\,\eta_H\right) > \hk^2\,\left(\hP\,\hT + \mu^2\,\hH^2\right)\,.
\label{eq:gdi}
\eq 
In the absence of Hall diffusion, the above criterion becomes
\bq
-\vp\,g\,\eta_A > \hk^2\,\hP\,\hT \,.
\label{eq:aC}
\eq
As the right hand side in the above equation is positive, the stability criterion implies that in the presence of a favourable shear flow gradient ($- \vp > 0$) ambipolar diffusion will drive the instability only if $g > 0$. Positive topological factor $g$ guarantees that the ambipolar diffusion will assist the shear flow in destabilising the waves. 

When ambipolar diffusion is negligible, Hall diffusion can as well drive the shear instability, provided
\bq
-\vp\,s\,\eta_H > \hk^2\,\left(\hO^2 + \mu^2\,\hH^2\right)\,.
\label{eq:hC}
\eq    
The above criterion is similar to Eq.~(24) of PW12a. Note that in PW12a, $s= 1$ as both the field and the wavevector is vertical whereas here we are dealing with the general field topology ($s \neq 1$). 

The instability in the Hall-ambipolar diffusion dominated regime will grow at the maximum rate
\begin{eqnarray}
\hs_0 = \gamma\, \frac{\left(\hP + \hT\right)-2\,\sqrt{\hP\,\hT + \mu^2\,\hH^2}}{\left(\hP - \hT\right)^2 -√± 4\,\mu^2\,\hH^2}
\,,
\label{eq:hamx}
\end{eqnarray}
which for the purely Hall ($\eta_A = \eta_O = 0$) case reduces to Eq.~(\ref{eq:hmax1}). In the absence of Hall diffusion the maximum growth rate becomes
\bq
\hs_0 = \frac{g\,\hA}{\left(\sqrt{\hP} + \sqrt{\hT}\right)^2}\,.
\label{eq:hdl1}
\eq
When $\hO = 0$ above equation reduces to  Eq.~(\ref{eq:amax}). 

Most unstable wavelength in the highly diffusive limit becomes
\bq
\hk_0^2 = \frac{2\,\hs_0^2}{\left[\alpha\,\left(g\,\hA + s\,\hH\right) - \left(\hP + \hT\right)\,\hs_0\right]}\,. 
\label{eq:kd}
\eq
For the purely Hall case and $\alpha = 1$ the preceding equation acquires a particularly simple form
\bq
\hk_0 = \sqrt{\frac{1}{2\,\Bz\,\hH}}\,.
\eq
In the purely ambipolar regime $\hk_0$ becomes, 
\bq
\hk_0 = \frac{1}{\left(1 + \mu\right)}\,\sqrt{
\frac{g}{\mu\,\hA}
}\,,
\eq
from where it is clear that the most unstable wavenumber is nonzero only when the topological parameter $g$ is nonzero. This is not surprising given that very existence of the ambipolar diffusion driven shear instability depends on the field geometry and obliqueness of the wave vector.  

\section{Discussion}
The solar photosphere is threaded by a kilogauss magnetic field concentrated in vertical flux tubes ($\mbox{radius} \sim 100-200\,\mbox{km}$) at intergranular boundaries \citep{H09}. Similar field strengths have also been observed in the quiet solar internetwork region \citep{SA10}, although less frequently ($\lesssim 40\,\%$). Outside these kilogauss patches, theinternetwork field is much weaker ($\sim 100\,\mbox{G}$). Flux tubes are often modelled as non-rotating cylindrical tubes with plasma and the magnetic pressure balance providing the required stability. In the present work, because we are interested in short radial wavelengths  we have approximated flux tubes by a planar sheet where the $x$ and $y$ coordinates locally correspond to the radial and azimuthal directions locally. 

Although the chromosphere has a limited extent in comparison to the corona, its net radiative loss $\sim 10^{7}\,\mbox{erg}\,\mbox{cm}^{-2}\,\mbox{s}^{-1}$ is $10$ times larger. Further, except for flares, most solar atmospheric heating occurs in the chromosphere \citep{A01, G01}. A strong correlation between the core emission of calcium K and H resonance lines and the quiet sun magnetic field \citep{S89} suggests that the origin of chromospheric heating ($\sim 10^7\,\mbox{ergs}\,\mbox{cm}^{-2}\,\mbox{s}^{-1}$) is magnetic. The magnetic-diffusion-driven shear instability proposed in the present work can provide a viable mechanism for the excess chromospheric heating as the crucial ingredients required to excite this instability -- shear flow and magnetic field -- are always present in the network--internetwork region. The attractive feature of this fast growing diffusive shear instability is that all wavelengths of fluctuations are likely to be excited as there is no cut-off wavelength. Thus, as we see from Fig.~(\ref{fig:DF2}), the network field below $1\,\mbox{Mm}$ is likely to be destabilised by this instability. In the internetwork elements, this instability may operate in the entire photosphere--chromosphere.

The only uncertainty involved is the lack of information about the scale of shear flow gradient which can not be resolved by current observations.  However, small whirlpools with size similar to terrestrial hurricanes [$\lesssim 0.5\,\mbox{Mm}$ with typical lifetime $\sim 5\,\mbox{min.}\,$ on the solar surface \cite{B08}] suggest the presence of such flows.  Long lasting large scale vortices at supergranular junctions with typical lifetime 
$\sim 1-2\,\mbox{h}$ with enhanced CaII emission have also been observed \citep{At09}. The swirl motion in the chromosphere has been recently detected by \cite{W09}. Clearly, observations suggest the ubiquitous presence of flow gradients in the photospheric-chromospheric plasma. Past numerical simulations also indicated the presence of vortex flows in intergranular lanes \citep{Z93, SN98}. The typical vorticity of a vortex is $\sim 6\times 10^{-3}\,\persec$ which corresponds to rotation period $\sim 35\,$ minutes \citep{B10}. Thus it would appear that the Hall instability does not have time to develop since the growth rate ($\avp / 2 = 3\times 10^{-3}\,\persec$) is very small. However, above vorticity value is limited by the upper limit in the vorticity resolution [$\sim 4\times 10^{-2}\,\persec$, \cite{B10}]. The numerical simulation gives much higher vorticity value ($\sim 0.1-0.2\,\persec$) in the photosphere-lower chromosphere (Fig.~31, \cite{SN98}). The growth rate corresponding to $\avp = 0.2\,\persec$ is one minute.  

For almost magnetosonic waves ($\kz \rightarrow 0$), the maximum growth rate of the Hall diffusion driven shear instability may become quite small.  The maximum growth rate in the ambipolar diffusion dominated middle and upper chromosphere will be  only one fifth of the Hall diffusion dominated case for the maximum $g = 0.25$ and $\mu = 0.5$. Therefore, vortex motions with typical lifetime $\gtrsim 15\,\mbox{min.}$ will be susceptible to the ambipolar diffusion driven shear instability. Since, vortex motions of various spatial and temporal scales are observed, it is likely that non--ideal MHD  effects will play an important role in exciting low frequency turbulence in the medium.    

\begin{figure}
%\vspace*{4cm}
 \includegraphics[scale=0.31 ]{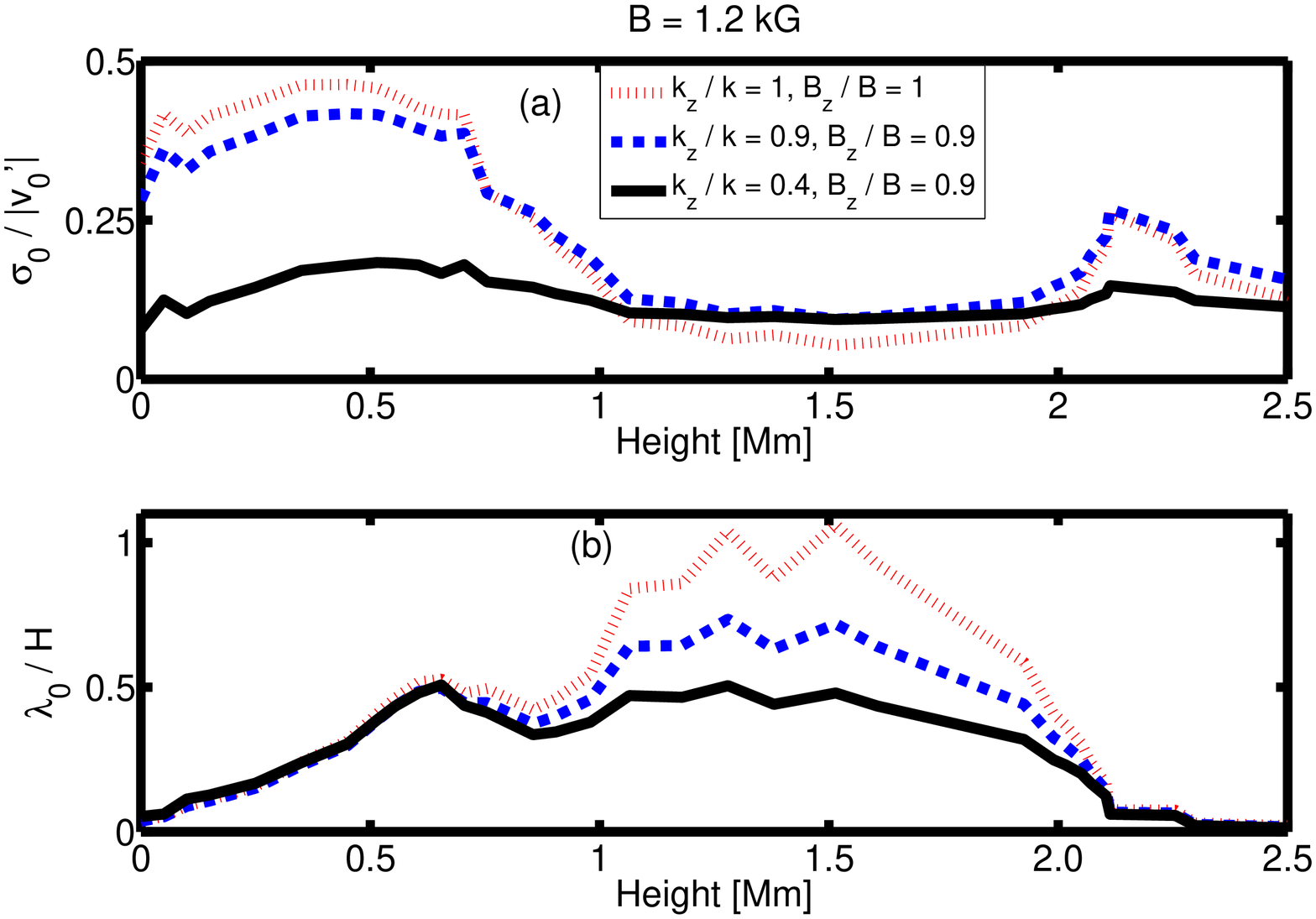}
\includegraphics[scale=0.31 ]{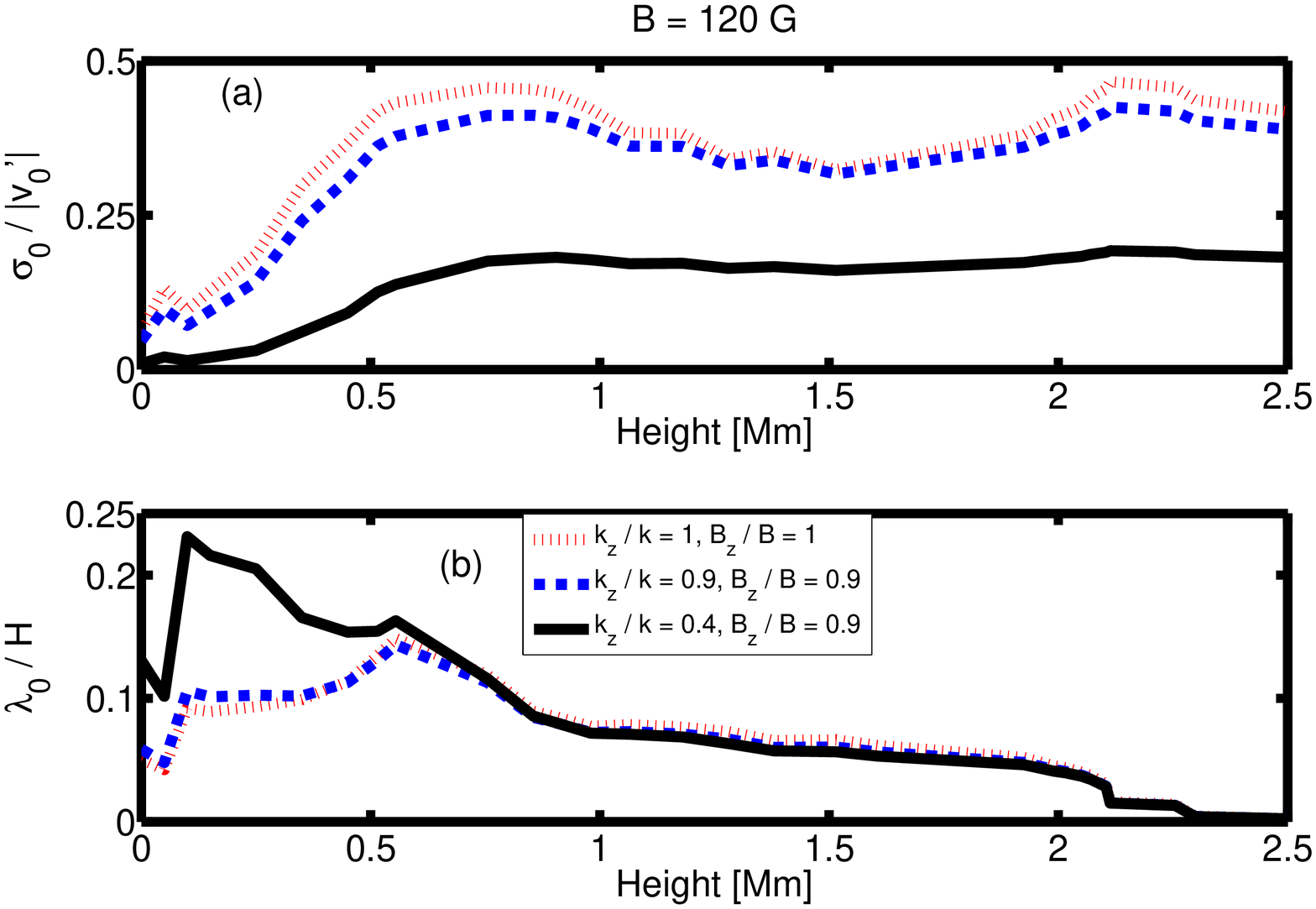}
\caption{The growth rate and most unstable wavelength is shown for $1.2\,\mbox{kG}$ [Fig.~\ref{fig:fr7}(a)] and $120\,\mbox{G}$ [Fig.~\ref{fig:fr7}(b)] fields. Following parameters have been used in the above figure: $\kz = 1\,,\Bz = 1$ (bold line), $\kz = \Bz = 0.9$ (dashed line) and $\kz = 0.4\,,\Bz = 0.9$ (dotted line).}
 \label{fig:fr7}  
\end{figure}

The \emph{non-ideal} MHD description of the photosphere-chromosphere provides several pathways through which shear energy can be channelled to the waves by magnetic field. For example in the excessively diffusive limit when $\dv / v \ll \dB / B$, diffusion in tandem with the shear flow can destabilise the network-internetwork field. For a purely vertical field and vertical wavevector this limit has been discussed in detail in PW12a.  In order to compare with PW12a, we briefly describe the effect of the field topology and wave orientation in the highly diffusive limit. Comparing Figs.~\ref{fig:DF2} and \ref{fig:fr7} we conclude that the maximum growth rate is similar in both cases. When $\Bz = \kz = 1$, the instability grows at a maximum rate whereas with decreasing $\kz$ or, $\Bz$ the growth rate diminishes.  

In Fig.~\ref{fig:fr7}(b) we plot most unstable wavelength against height for  the same parameters as in Fig.~\ref{fig:fr7}(a). The wavelength is normalized against scale height calculated self--consistently using model C, VAL81.   
For both kG and weaker fields $\lambda_0$ fits well within a scale height and thus the instability will grow at a maximum rate in the entire photosphere--chromosphere. However, it is only in the photosphere and lower chromosphere ($\lesssim 1\,\mbox{Mm}$) where the instability will grow at a maximum rate for a kG field. In the middle and upper chromosphere the growth rate tapers off and becomes one eighth of the shear frequency. Therefore, in the strong field region, diffusive instability will be efficient in destabilizing a magnetic element in the photosphere and lower chromosphere. When the field is weak ($\sim 100\,\mbox{G}$), the instability can operate in the entire photosphere-chromosphere at a maximum rate.   

How does a nonlinear saturated state of the diffusive instability will look like? The answer to this can be given only by numerical simulations. However, if the nonlinear results of the protoplanetary discs and star forming regions are any guide then this instability should be quite efficient in exciting the low frequency turbulence and heating of the plasma. In fact the interplay between the vortex flow and magnetic diffusion could be responsible for the entire energy budget of the solar atmosphere.  

The energetic events such as hard x-ray emissions ($\sim 10^{26}\,\mbox{erg}\,\mbox{s}^{-1}$) are believed to be due to the presence of energetic electrons with energies above ($\sim 20\,\mbox{keV}$) in solar flares. Although, the exact mechanism of the initiation and triggering of solar flares is not yet known it is widely believed that the flares and the associated eruptions may occur due to magnetic reconnection, i.e., the rapid dissipation of electric currents near the magnetic null points. The ensuing relaxation of the sheared magnetic field topology can give rise to the large--scale \alf waves which may transport the energy to the chromosphere \citep{FH08}. Thus the development of turbulence in the chromosphere via reflection and mode conversion       
Of the \alf wave may lead to the cascade of energy to the short wavelengths. The stochastic acceleration of the electron by a turbulent wave spectrum produces a high energy spectrum \citep{FH08}.  This model of electron acceleration crucially depends  on the \alf wave propagation along the  field line to the chromosphere. However, as we have noted in the introduction, the concept of well defined flux tubes in  a highly magnetic diffusion dominated medium is unclear. However, the electron acceleration in the chromosphere may indeed take place as envisioned by   the Fletcher-Hudson Model due to magnetic diffusion driven turbulence.  Since magnetic diffusion could be an important agent in driving the diffusive shear instability, this could easily lead to the low frequency turbulence in the medium. However, further work is needed in this direction to support this hypothesis.

\section{summary}
The granular motion is responsible for the generation of low frequency \alf waves in the predominantly neutral photosphere. Unlike the high-frequency MHD counterparts which undergo damping \citep{VPP07, VPPD07}, low frequency \alf waves generally propagate undamped in the medium. In the presence of a shear flow gradient, the solar photosphere-chromosphere region can become unstable due to non-ideal MHD effects. Depending on the shear scale, such an instability can be excited at all wavelengths. The presence of a favourable magnetic field topology will facilitate the transfer of shear energy to the magnetic fluctuations. The instability in the solar atmosphere is due to the presence of two unrelated physical processes   (a) \emph{nonideal} MHD effects and, (b) a shear flow.          

The following is an itemized summary of the present work.

1. The magnetic diffusion in the presence of shear flow makes the solar atmosphere susceptible to the diffusive shear instability. The instability depends on the sign of the shear gradient and on the local magnetic field topology.

2. A purely vertical magnetic field and vertical wavevector along with the shear flow is fundamental to this Hall diffusion driven instability.

3. Only when the field has a non-vertical component and when waves are propagating obliquely, both Hall and ambipolar diffusion together can assist this instability.   

4. The maximum growth rate of the instability which is proportional to the absolute value of the shear gradient occurs when both the field and the wavevector is vertical. 

5. The e-folding time of the magnetic diffusion driven shear instability is very short ($\sim 20\,\mbox{s}$ for $\vp = 0.1\,\persec$)  suggesting that the flux tubes are likely to be susceptible to this fast growing instability.    

\section*{Acknowledgments}
{The financial support of Australian Research Council through grants DP0881066  and DP120101792 is gratefully acknowledged. This research has made use of NASAs Astrophysics Data System.}

\appendix

\section[]{ }

Often, the ion√±-neutral collision frequency is calculated by assuming a collision cross√±-section $\sigma = 10^{-15}\,\mbox{cm}^{2}$ \citep{K104}. This cross--section is even smaller than that of the $H^{+} + H$ collision frequency. With this it becomes clear if we recall that proton√±-neutral hydrogen cross√±-section is well described by the following power law at low energies \citep{G05}
\bq
\sigma(E) = \sigma(E_1)\,\left(\frac{E}{E_1}\right)^p\,
\label{eq:CS1}
\eq       
where $E_1 = 0.01\,\mbox{eV}$ and $\sigma(E_1) = 1.65 \times 10^{-14}\,\mbox{cm}^2\,\left(590\,\mbox{amu}\right)$. In the solar atmosphere, $E > E_1$ and $p = - 1/8$. For Model C, VAL81, $\sigma \approx 10^{-14}\,\mbox{cm}^2$. Thus, if we assume that the ion-neutral collision is solely due to the ionised and neutral hydrogen, the ion-neutral collision frequency will be an order of magnitude higher than when $\sigma \approx 10^{-15}\,\mbox{cm}^2$.  

However, the lower solar atmosphere is populated mainly by the metallic ions and atomic hydrogen and thus, the collision cross-section given by Eq.~(\ref{eq:CS1}) underestimates the collision frequency. Further, the quantum effects could also become important. Thus, it is important to find an analytical expression for the momentum transfer cross√±-section. At low energies ($\leq 10^4\,\mbox{K}$), the elastic cross√±-section can be well approximated by the Massey√±-Mohr cross section \citep{K09}
\bq
\sigma_{MM} = 5\times 10^{11}\,\left(\frac{C_6}{v}\right)^{2/5}\,\mbox{cm}^2\,. 
\eq      
Here $C_6$ is the dipole√±-dipole coefficient in the interaction potential which can be expressed in terms of polarisabilities $\alpha$ and ionisation energies $E$ of two interacting atoms \citep{F61}
\bq
C_6 = 1.5\,\alpha_1\,\alpha_2\,\frac{E_1\,E_2}{E_1 + E_2}\,.
\eq
The value of $C_6$ varies between $1.26 \times 10^{-60}\,\mbox{erg} -\mbox{cm}^3$ for He to $2200 \times 10^{-60}\,\mbox{erg}-\mbox{cm}^3$ for Cs \citep{F61}. Therefore, we assume $C_6 = 100 \times 10^{-60}\,\mbox{erg}-\mbox{cm}^3$. Since the cross√±-section is proportional to $C_6^{0.4}$, the assumed value of $C_6$ will not significantly affect the collision cross--section. The collision rate is
\bq
<\sigma_{MM} v>_{in} = 3.15\times 10^{12}\,v^{3/5}\,\mbox{cm}^3\,\mbox{s}^{-1}\,
\eq
which can also be written as
\bq
 <\sigma_{MM} v>_{in} = 7.91\times 10^{-10}\,T^{0.3}\,\mbox{cm}^3\,\mbox{s}^{-1}\,.
\eq
Here we have assumed $v = \sqrt{k_B\,T / m_i}$. The resulting value of $\nu_{in}$ is 
\bq
\nu_{in} = \frac{<\sigma_{MM} v>_{in}}{m_i + m_n}\rho_n\,,
\eq
which is an order of magnitude smaller than has been assumed by \cite{VPPD07} owing to their arbitrary mass-√±scaling of the cross√±-section.

The electron√±-neutral collision rate is assumed as
\bq       
<\sigma\, v>_{en} = 10^{-15}\,\left(\frac{128\,k_B\,T}{9\,\pi\,m_e}\right) \equiv 8.28\times\,10^{-10}\,\sqrt{T}\,\mbox{cm}^3\,\mbox{s}^{-1}\,. 
\eq 
For similar reasons, the electron√±-neutral collision frequency ($\nu_{en} = n_n\,<\sigma\, v>_{en}$) in the present case is an order of magnitude smaller than given by \cite {VPPD07, P08b}. 
\end{document}